\documentclass[aps,prl,twocolumn,showpacs,superscriptaddress,showkeys,bm,amsmath,amssymb,longbibliography]{revtex4-1}

\usepackage{xcolor}
\usepackage{graphicx}
\usepackage{multirow}

\begin{document}
\title{Search for Sub-eV Sterile Neutrino at RENO}

\newcommand{\CNU}{\affiliation{Institute for Universe and Elementary Particles, Chonnam National University, Gwangju 61186, Korea}}
\newcommand{\DSU}{\affiliation{Institute for High Energy Physics, Dongshin University, Naju 58245, Korea}}
\newcommand{\GIST}{\affiliation{GIST College, Gwangju Institute of Science and Technology, Gwangju 61005, Korea}}
\newcommand{\IBS}{\affiliation{Institute for Basic Science, Daejeon 34047, Korea}}
\newcommand{\KAIST}{\affiliation{Department of Physics, KAIST, Daejeon 34141, Korea}}
\newcommand{\KNU}{\affiliation{Department of Physics, Kyungpook National University, Daegu 41566, Korea}}
\newcommand{\SNU}{\affiliation{Department of Physics and Astronomy, Seoul National University, Seoul 08826, Korea}}
\newcommand{\SYU}{\affiliation{Department of Fire Safety, Seoyeong University, Gwangju 61268, Korea}}
\newcommand{\SKKU}{\affiliation{Department of Physics, Sungkyunkwan University, Suwon 16419, Korea}}

\author{J.~H.~Choi}\DSU
\author{H.~I.~Jang}\SYU
\author{J.~S.~Jang}\GIST
\author{S.~H.~Jeon}\SKKU
\author{K.~K.~Joo}\CNU
\author{K.~Ju}\KAIST
\author{D.~E.~Jung}\SKKU
\author{J.~G.~Kim}\SKKU
\author{J.~H.~Kim}\SKKU
\author{J.~Y.~Kim}\CNU
\author{S.~B.~Kim}\SKKU
\author{S.~Y.~Kim}\SNU
\author{W.~Kim}\KNU
\author{E.~Kwon}\SKKU
\author{D.~H.~Lee}\SKKU
\author{H.~G.~Lee}\SNU
\author{I.~T.~Lim}\CNU
\author{D.~H.~Moon}\CNU
\author{M.~Y.~Pac}\DSU
\author{H.~Seo}\SNU
\author{J.~W.~Seo}\SKKU
\author{C.~D.~Shin}\CNU
\author{B.~S.~Yang}\IBS
\author{J.~Yoo}\IBS\KAIST
\author{S.~G. Yoon}\KAIST
\author{I.~S.~Yeo}\CNU
\author{I.~Yu}\SKKU

\collaboration{The RENO Collaboration}
\noaffiliation
\date{\today}

\begin{abstract}
We report a search result for a light sterile neutrino oscillation with roughly 2\,200 live days of data in the RENO experiment.
The search is performed by electron antineutrino ($\overline{\nu}_e$) disappearance taking place between six 2.8\,GW$_{\text{th}}$ reactors and two identical detectors located at 294\,m (near) and 1383 \,m (far) from the center of reactor array.
A spectral comparison between near and far detectors can explore reactor $\overline{\nu}_e$ oscillations to a light sterile neutrino. An observed spectral difference
is found to be consistent with that of the three-flavor oscillation model.
This yields limits on $\sin^{2} 2\theta_{14}$ in the $10^{-4} \lesssim |\Delta m_{41}^2| \lesssim 0.5$\,eV$^2$ region, free from reactor $\overline{\nu}_e$ flux and spectrum uncertainties.
The RENO result provides the most stringent limits on sterile neutrino mixing at $|\Delta m^2_{41}| \lesssim 0.002$\,eV$^2$ using the $\overline{\nu}_e$ disappearance channel.
\end{abstract}
\pacs{13.15.+g,14.60.Pq, 14.60.St, 28.50.Hw, 29.40.Mc}

\keywords{sterile neutrino, RENO, reactor neutrino, neutrino oscillation}

\maketitle


There remain unknown properties of neutrino even with impressive progress in neutrino physics.
 The number of neutrino flavors is not firmly determined yet.
 Almost all the experimental results indicate that the number of light neutrino species is consistent with only three-flavors.
However, some of experimental results may not be explained by the three active flavor neutrino hypothesis \cite{Athanassopoulos:1996jb,Hampel:1997fc,Abdurashitov:2009tn,Mention:2011rk,Aguilar-Arevalo:2013pmq,Giunti:2010zu} and
suggest an additional flavor of neutrino with a mass around 1\,eV, called as sterile neutrino because of no interaction with ordinary particles \cite{Pontecorvo:1967fh}.

 Short baseline (SBL) experiments with their detectors located at a few tens of meters from a reactor are carried out or proposed to search for a sterile neutrino  \cite{Ko:2016owz,Alekseev:2018efk}.
 The SBN project at FNAL \cite{Antonello:2015lea} and the JSNS$^{2}$ experiment at J-PARC \cite{Ajimura:2017fld} are going to search for sterile neutrino oscillation using accelerator beams.

An interesting motivation for investigating a sub-eV sterile neutrino comes from cosmological data.
 Recent Planck data \cite{Ade:2015xua} seems to rule out
  an additional neutrino species with a mass near 1\,eV assuming full thermalization in the early Universe. 
  However, sterile neutrinos have played an important role in explaining the dark radiation excess and the preference for a hot dark matter component with mass in the sub-eV range \cite{Gariazzo:2013gua}.

Reactor experiments of Daya Bay, Double Chooz and RENO can search for lighter sterile neutrinos with multiple identical detectors and baselines of $\sim$1\,km \cite{Palazzo:2013bsa}. 
The Daya Bay Collaboration reported the sub-eV sterile neutrino search result \cite{An:2014bik,An:2016luf}.

This Letter reports a search for a light sterile neutrino based on the 3+1 neutrino hypothesis using more than one million reactor $\overline{\nu}_e$ interactions in the RENO experiment.
 According to this hypothesis, the survival probability for $\overline{\nu}_e$ with an energy $E$ and a distance $L$ is approximately given by  \citep{Palazzo:2013bsa}	
\begin{eqnarray}
  P_{\overline{\nu}_e \rightarrow \overline{\nu}_e}
  \approx 1
  &-& \sin^{2}2\theta_{13}\sin^{2}\Delta _{13}
    \nonumber \\\
     &-& \sin^{2}2\theta_{14}\sin^{2}\Delta _{41}
  , \quad \quad
\label{eq:psurv-4nu-sim}
\end{eqnarray}
where $\Delta_{ij} \equiv 1.267 \Delta m^2_{ij} L/E$, 
 $\Delta m_{ij}^2 \equiv m_i^2 – m_j^2$ is the mass-squared difference between the mass eigenstates.
This indicates the sterile neutrino oscillation with a mixing angle $\theta_{14}$ introduces an additional spectral distortion by a squared mass difference $|\Delta m_{41}^2|$. Thus these oscillation parameters can be explored by a model independent spectral comparison of the reactor $\overline{\nu}_e$ disappearance between near and far detectors. 
In this Letter, RENO presents a result of the light sterile neutrino search in its sensitive region of $|\Delta m^2_{41}|  \lesssim$ 0.5\,eV$^2$.

The RENO experiment uses two identical near and far detectors located at 294 and 1383\,m, respectively, from the center of six reactor cores at the Hanbit Nuclear Power Plant Complex in Yonggwang.
The near (far) underground detector has 120\,m (450\,m) of water equivalent overburden.
Six pressurized water reactors, each with maximum thermal output of 2.8\,GW$_{\text{th}}$, are situated in a linear array spanning 1.3\,km with equal spacings.
The reactor flux-weighted baseline is 410.6\,m for the near detector and 1445.7\,m for the far detector, respectively.
The baseline distances between the detectors and reactors are measured to an accuracy or better than 10\,cm using GPS and total station.
      	
Each RENO detector consists of a main inner detector,
 filled with 16 tons of 0.1\,\% gadolinium (Gd) loaded liquid scintillator, and an outer veto detector. 
A reactor $\overline{\nu}_e$ is detected through the inverse beta decay (IBD) reaction, $\overline{\nu}_e + p \rightarrow e^+ + n $.
Backgrounds are efficiently removed by time coincidence between a prompt signal and a delayed signal from neutron capture on Gd. 
The prompt signal releases energy of 1.02\,MeV as two $\gamma$ rays from the positron annihilation in addition to the positron kinetic energy. The delayed signal produces several $\gamma$ rays with the total energy of $\sim$8\,MeV. 
The details of the RENO detector are described in Refs. \cite{RENO:2015ksa,Park:2012dv,Park:2013nsa,Ma:2009aw,Bak:2018ydk}.

Due to various baselines between two detectors and six reactor cores, ranging from three hundred meters to nearly 1.5 kilo-meters as shown in Table \ref{tab:baseline}, this search is sensitive to mixing between active and sterile neutrinos in the region of $10^{-4} \lesssim |\Delta m_{41}^2| \lesssim 0.5$\,eV$^2$. These mixing parameters can produce an additional modulation in energy with a frequency different from the active neutrino oscillation.

\begin{table}[thb]
  \centering
  \caption{Baselines of near and far detectors from the six reactor cores.  
    \label{tab:baseline}}
  \begin{ruledtabular}
    \begin{tabular}[c]{lcccccc} 
      \multirow{2}*{Detectors} 
      & \multicolumn{6}{c}{Baselines (m)} \\
       \cline{2-7}  
      & R1 & R2 & R3 & R4 & R5 & R6 \\
      \hline
        Near & 660 & 445 & 302  &  339 & 520  &     746 \\
        Far  & 1564 & 1461 & 1398 &  1380  &  1409 &    1483\\ 
    \end{tabular}
  \end{ruledtabular}
\end{table}

This analysis uses roughly 2\,200 live days of data taken in the period between August 2011 and February 2018.
Applying the IBD selection criteria yields 850\,666 (103\,212) $\overline{\nu}_e$ candidate events with the energy of prompt event ($E_p$) between 1.2 and 8.0\,MeV in the near (far) detector.
The background fraction for the near (far) detector is 2.0\,\% (4.8\,\%).
The $E_p$ resolution in the range of 1 to 8\,MeV is 8 to 3\,\%.
 A detailed description of IBD event selection criteria and their systematic uncertainties can be found in Refs. \cite{RENO:2015ksa,Seo:2016uom,Bak:2018ydk}. 

The uncertainty in the absolute energy scale is estimated to be 1.0\,\% \cite{Seo:2016uom}.
This sterile neutrino search based on the relative measurement of spectra at two identical detectors is almost insensitive to the uncertainty. On the other hand, the $E_p$ difference between
 the near and far detectors contributes to the uncertainty associated with this analysis.
The relative energy scale difference is estimated by comparing near and far spectra of calibration data and is found to be less than 0.15\% \cite{Seo:2016uom}.
The finite sizes of the reactor cores and the antineutrino detectors, relevant to a search in the region of $|\Delta m^2| \sim$ 1\,eV$^2$, make a negligible effect on the sterile neutrino search in the RENO’s sensitive region of $|\Delta m^2| \lesssim 0.5$\,eV$^2$.
The expected rates and spectra of reactor $\overline{\nu}_e$ are calculated for the duration of physics data taking by taking into account the varying thermal powers, fission fractions of four fuel isotopes, energy release per fission, fission spectra, IBD cross sections, and detector response \cite{Seo:2016uom}.

\begin{figure}[tb!]
\centering
\includegraphics[width=0.98\linewidth]{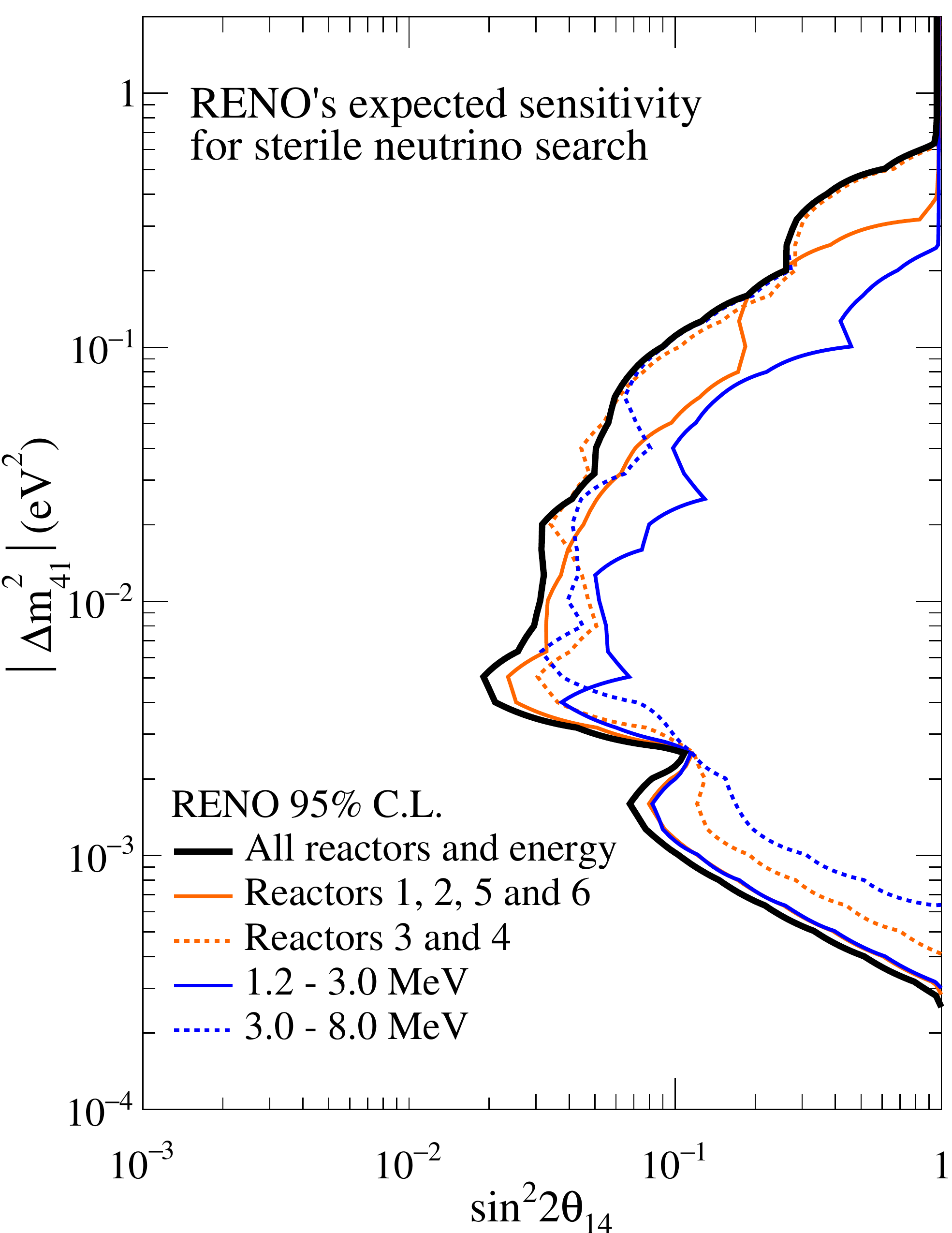}
\caption{(Colors online) Expected 95\,\% C.L. exclusion contours from sterile neutrino searches.
 The black solid contour represents an expected limit on $\overline{\nu}_e$ disappearance using the RENO’s 2200 days of data.
  The red solid (dotted) contour represents an exclusion sensitivity originating from a relatively long (short) baseline search. The blue solid (dotted) contour represents an exclusion sensitivity coming from a search in the 1.2 – 3.0\,MeV (3.0 – 8.0\,MeV) region.
} 
\label{fig:ASMV_MC_sen}
\end{figure}

The RENO’s multiple reactors provide various baselines between the near and far detectors for exploring a sterile neutrino oscillation in a wide range of $|\Delta m_{41}^2|$ values.
With the various baselines and energies of reactor neutrino, a sensitivity study for an excluded parameter region is performed using an Asimov Monte Carlo method \cite{Cowan:2010js}.
The sample is generated without statistical or systematic fluctuations assuming the three-neutrino hypothesis.
Figure \ref{fig:ASMV_MC_sen} shows an Asimov expected exclusion contour obtained from a
search for a sterile neutrino oscillation by a far-to-near ratio method which is described later.
In the $10^{-4} <|\Delta m^2_{41}|< 0.5$\,eV$^2$ region,
a relative spectral distortion between the two detectors occurs and obtains a search sensitivity.
The dip structure at 0.003\,eV$^2$ is caused by a degenerate oscillation effect due to $\theta_{13}$ and $\theta_{14}$.
In the $|\Delta m^2_{41}| < 10^{-4}$\,eV$^2$ region,
an oscillation length becomes longer than the baseline distance between the  two detectors and loses a search sensitivity.
The sensitivity in the $0.01 \lesssim |\Delta m_{41}^2| \lesssim 0.5$\,eV$^2$ ($|\Delta m_{41}^2| \lesssim 0.01$\,eV$^2$) region comes from the spectral comparison at relatively short (long) baselines between the two detectors or from the prompt energy above (below) 3\,MeV.
In the $|\Delta m^2_{41}| \gtrsim 0.5$\,eV$^2$ region, the far-to-near ratio method is unable to exclude any parameter region because of no relative spectral distortion between the two detectors.
A rapid oscillation takes place before the near detector in the large $|\Delta m^2_{41}|$ region and generates no spectral distortion between the two detectors.
However, comparison of their event rates becomes sensitive to exclude oscillation parameters.

\begin{figure}[tb!]
\centering
\includegraphics[width=0.48\textwidth]{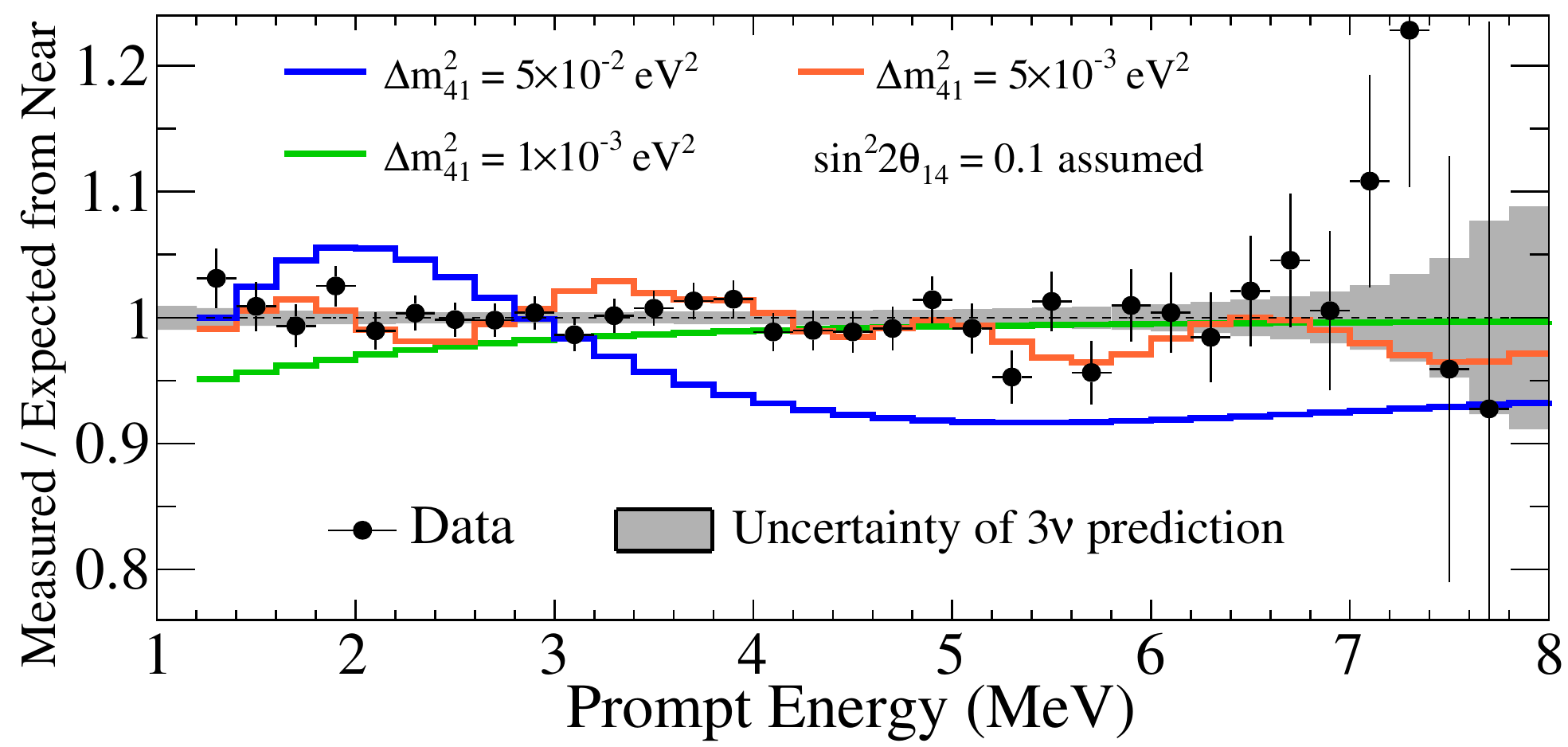}
\caption{
(Colors online) Prompt energy spectra observed
at far detector divided by the 3 neutrino best-fit prediction from the near detector spectrum \cite{Tanabashi:2018oca}.
 The gray band represents the statistical uncertainty of the near data and all the systematic uncertainties. 
 Predictions with $\sin^2 2\theta_{14}$ = 0.1 and three $|\Delta m^2_{41}|$ representative values are also shown as the blue, red and green curves.}
\label{fig:ex}
\end{figure}

This sterile neutrino search is based on comparison of observed spectra with two identical detectors having different baselines, and thus independent of a reactor $\overline{\nu}_e$ flux and spectrum model.
A sterile neutrino oscillation causes $\overline{\nu}_e$ disappearance according to Eq. (\ref{eq:psurv-4nu-sim}) and produces relative spectral distortion between the near and far detectors.
Figure \ref{fig:ex} shows the ratio of
the observed prompt energy spectrum at far detector 
and the 3 neutrino best-fit prediction from the near detector spectrum \cite{[{}][{ and 2019 update}]Tanabashi:2018oca}.
The 3+1 neutrino oscillation predictions are also shown for $\sin^2 2\theta_{14} = 0.1$ and three $|\Delta m_{41}^2|$ values. The comparison between data and predictions demonstrates RENO’s sensitivity of $|\Delta m_{41}^2| \lesssim 0.5$\,eV$^2$ in exploring a sterile neutrino oscillation.
Due to the discrepancy of observed flux and spectra from the reactor $\overline{\nu}_e$ model prediction,
this analysis employs the relative spectral distortion between identical near and far detectors.
Moreover, the spectral ratio comparison
cancels out common systematic uncertainties between the two identical detectors.
 The active and sterile oscillation parameters are determined by a fit to the measured far-to-near ratio of IBD prompt spectra in the same manner as the previous three-neutrino oscillation analysis \cite{Bak:2018ydk}.
To find the best fit, a $\chi^2$ with pull parameter terms of systematic uncertainties is constructed using the spectral ratio measurement and is minimized by varying the oscillation parameters and pull parameters as described in Ref. \cite{Bak:2018ydk}:
\begin{eqnarray}
 \chi^2  & = & \sum_{i=1}^{N_{bins}} \frac{(O_i^{F/N} - T_i^{F/N})^2 }{U_i^{F/N}}
                          + \sum_{d=N, F} \left( \frac{b^{d}}{\sigma_{bkg}^{d}} \right)^2 
 \nonumber       \\
&&                        + \sum_{r=1}^{6} \left( \frac{f_r}{\sigma_{flux}^r} \right)^2 
                          + \left( \frac{\epsilon}{\sigma_{eff}} \right)^2
                          + \left( \frac{e}{\sigma_{scale}} \right)^2 ,
 \label{eq:chi_sq}                         
\end{eqnarray} 
where $O_i^{F/N}$ and $T_i^{F/N}$ are the observed and expected far-to-near ratio of IBD events in the $i$-th $E_p$ bin,
$U_i^{F/N}$ is the statistical uncertainty of $O_i^{F/N}$, and
$O_i^{F/N}$ is the ratio of the spectra after background substraction as Ref. \cite{Bak:2018ydk}.
The expected far-to-near ratio is calculated using reactor and detector information including pull parameters ($b^d$, $f_r$, $\epsilon$, and $e$).
The systematic uncertainty sources are embedded by these pull parameters  with associated systematic uncertainties ($\sigma_{bkg}^d$, $\sigma_{flux}^r$, $\sigma_{eff}$, and $\sigma_{scale}$).
 The details of pull terms and systematic uncertainties are described in Ref. \cite{Bak:2018ydk}.
The $\chi^2$ is minimized with respect to the pull parameters and the oscillation parameters.

The oscillation parameters of $\theta_{14}$, $\theta_{13}$ and $|\Delta m_{41}^2|$ are set as free.
The rest of variables are constrained with other measurements:
 $\sin^2 2 \theta_{12}$ = $0.846 \pm 0.021$, $\Delta m^2_{21} = (7.53 \pm 0.18) \times 10^{-5}\,\text{eV}^2$ and $|\Delta m^2_{32}| = (2.444 \pm 0.034) \times 10^{-3}\,\text{eV}^2$ \cite{Tanabashi:2018oca}.
However, the parameters of $\theta_{12}$ and $\Delta m^2_{21}$ are fixed because of their negligible effect on $\chi^2$. The parameter $\Delta m^2_{31}$ only is constrained by a pull term in the $\chi^2$.
The normal mass ordering is assumed for both $\Delta m^2_{31}$ and $\Delta m^2_{41}$.

The minimum $\chi^2$ value for the 3+1 neutrino hypothesis is $\chi^2_{4\nu}/$NDF = 46.4/65, where NDF is the number of degrees of freedom.
The value for the three-neutrino model with unconstrained $|\Delta m_{31}^2|$ is $\chi_{3\nu}^2$/NDF = 47.8/66 .
The distribution of $\chi^2$ difference between the two hypotheses,
$\Delta \chi^2 = \chi_{3\nu}^2 – \chi_{4\nu}^2$, is obtained from a number of simulated experiments with a statistical variation and their $\chi^2$ fits with systematic uncertainties taken into account.
The \textit{p}-value corresponding to the $\Delta \chi^2$ value is obtained to be 0.87 for $\Delta \chi^2$ = 1.4.
This indicates the data are found to be consistent with the 3 neutrino model and show no significant evidence for a sterile neutrino oscillation.

\begin{figure}[bt!]
\centering
\includegraphics[width=0.98\linewidth]{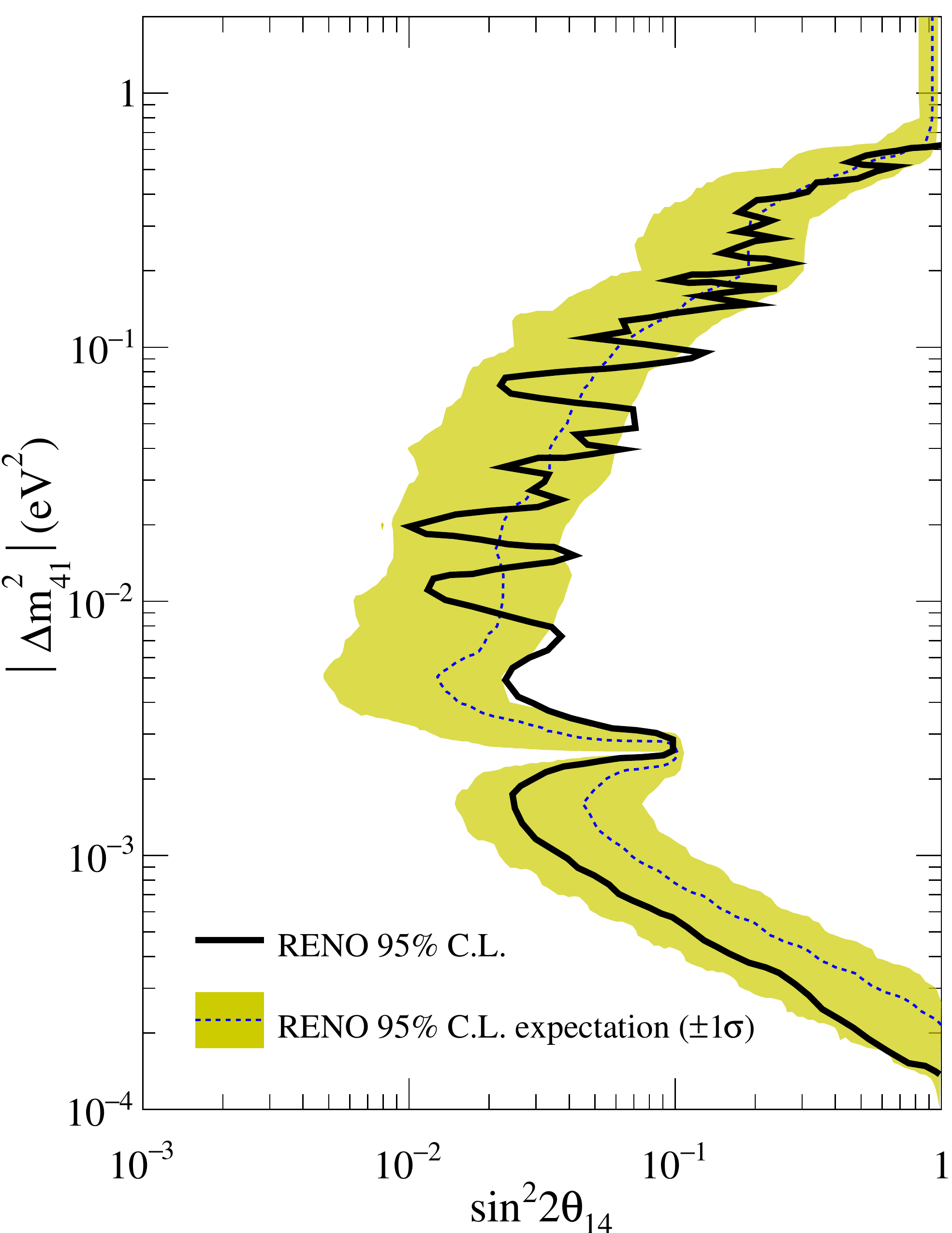}
\caption{(Colors online) 
RENO’s 95\,\% C.L. exclusion contour for the sterile neutrino oscillation parameters of $\sin^2 2\theta_{14}$ and $|\Delta m^2_{41}|$. 
The black solid contour represents an excluded region obtained from spectral distortion between near and far detectors.
 The green shaded band represents expected 1$\sigma$ exclusion contours due to a statistical fluctuation. 
 The blue dotted contour represents its median.
  The parameter region in the right side of the contours is excluded.
} 
\label{fig:Expect}
\end{figure}

Exclusion limits in a parameter space of $\sin^2 2\theta_{14}$ and $|\Delta m^2_{41}|$ are set on sterile neutrino oscillation by a standard $\Delta\chi^2$ method \cite{Tanabashi:2018oca}.
For each parameter set of $\sin^2 2\theta_{14}$ and $|\Delta m^2_{41}|$, $\Delta \chi^2 = \chi^2 - \chi^2_{min}$ is obtained, where $\chi^2_{min}$ is the minimum $\chi^2$ out of all possible parameter sets. The values of $\theta_{13}$ and $|\Delta m^2_{31}|$ are determined by the $\chi^2_{min}$. The parameter sets of $\sin^2 2\theta_{14}$ and $|\Delta m^2_{41}|$ are excluded at 95\,\% confidence level if $\Delta \chi^2$ is greater than 5.99 \cite{Tanabashi:2018oca}.
Figure \ref{fig:Expect} shows an exclusion contour obtained from the RENO data.
We repeat obtaining exclusion contours using the Gaussian CL$_s$ method \cite{Read:2002hq, Qian:2014nha}.
For each set of $\sin^2 2\theta_{14}$ and $|\Delta m^2_{41}|$, this method calculates \textit{p}-values for the three-neutrino and 3+1 neutrino hypotheses and determines a CL$_s$ value from them.
 A 95\% C.L. exclusion region is obtained by requiring a condition of CL$_s$ $\leq$ 0.05.
The $\Delta\chi^2$ and Gaussian CL$_s$ methods obtain 95\,\% C.L. contours of negligible difference within a statistical fluctuation. 

In order to understand the validity of the data analysis, a number of pseudo-experiments are generated within statistical fluctuation and without the sterile neutrino hypothesis.
 Exclusion contours for the pseudo-experiments
are obtained by the same $\Delta \chi^2$ method as described above, by taking into account the systematic uncertainties.
 Figure \ref{fig:Expect}
 also shows an expected 1$\sigma$ band of 95\,\% C.L. exclusion contours due to a statistical fluctuation and its median.
 The RENO's obtained exclusion contour is mostly contained in the 1$\sigma$ band.

The fluctuating behavior of the obtained exclusion contour in the region of $|\Delta m^2_{41}| \gtrsim 0.002$\,eV$^2$ comes from the finite size of the data sample.
 In the $|\Delta m^2_{41}| \lesssim 0.002$\,eV$^2$ region, the spectral distortion appears in the low energy range and gradually disappears.
The data exclude a larger range of $\sin^2 2\theta_{14}$ values than the Asimov prediction in this $|\Delta m^2_{41}|$ region.
The spectral deviation from the 3 neutrino prediction at low energy happens to be minimal and obtains a more excluded region than the most probable expectation.
According to pseudo-experiments, such an exclusion contour away from the expectation is estimated to have a probability of roughly 20\,\%.
A dip structure at $|\Delta m^2_{41}| \sim 0.003$\,eV$^2$ as found in the Asimov study is observed due to 
an oscillation degeneracy of $\theta_{13}$ and $\theta_{14}$.
  In the $|\Delta m^2_{41}| \gtrsim 0.5$\,eV$^2$ region, the spectral distortion due to the sterile neutrino oscillation is averaged out before the near detector and a search sensitivity is lost.

The limit of $\sin^2 2\theta_{14}$ is mostly determined
by a statistical uncertainty while the systematic uncertainties become considerable in the $|\Delta m^2_{41}| \lesssim 0.06$\,eV$^2$.
 The uncertainty of background ($\sigma_{bkg}^d$) is a dominant systematic source in the $0.003 \lesssim |\Delta m^2_{41}| \lesssim 0.06$\,eV$^2$ region,
 and the energy-scale uncertainty ($\sigma_{scale}$) is a major limiting factor in the $|\Delta m^2_{41}| \lesssim 0.008$\,eV$^2$ region.
  The uncertainties of flux ($\sigma_{flux}^r$)  and detection  efficiency ($\sigma_{eff}$) have negligible effect on this analysis.

\begin{figure}[bt!]
\centering
\includegraphics[width=0.98\linewidth]{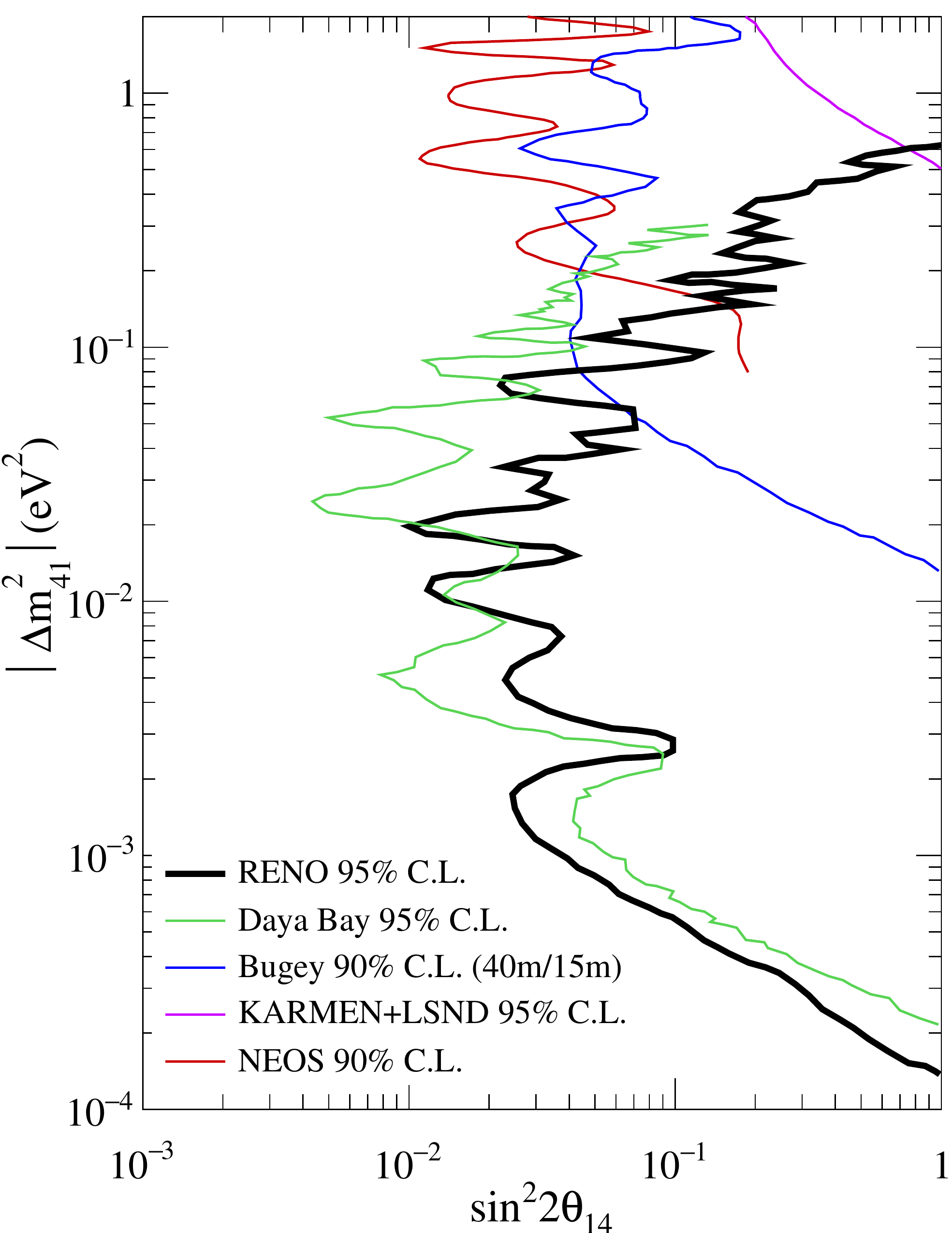}
\caption{(Color online) Comparison of the exclusion limits.
The right side of each contour shows excluded region.
The black solid line represents the 95\,\% C.L. exclusion contour using spectral distortion between near and far spectra. 
 For the comparison, Daya Bay's \citep{An:2016luf} 95\,\% C.L. (green), Bugey's \citep{Declais:1994su} 90\,\% C.L.(blue), KARMEN+LSND \cite{Conrad:2011ce} 95\,\% C.L.(magenta)  and NEOS's \citep{Ko:2016owz} 90\,\% C.L.(brown) limits on $\overline{\nu}_e$ disappearance are also shown.}
\label{fig:cont}
\end{figure}

Figure \ref{fig:cont} shows exclusion contours obtained from the RENO data and other experiments.
 The RENO spectral comparison between the near and far detectors yields stringent limits on $\sin^2 2\theta_{14}$ in the 10$^{-4} \lesssim |\Delta m^2_{41}| \lesssim 0.5$\,eV$^2$ region, while SBL reactor neutrino experiments are sensitive to the $|\Delta m^2_{41}| \gtrsim 0.01$\,eV$^2$ region.
 RENO's longer baselines than the SBL experiments allows sensitivity to search for lighter sterile neutrino mixing.
Combining the RENO result with those of other experiments can improve the sterile neutrino search sensitivity. More accurate SBL reactor and accelerator neutrino experiments are desirable in order to probe the $|\Delta m^2_{41}|$ larger than 0.5\,eV$^2$.

In summary, RENO reports results from a search for a sub-eV sterile neutrino oscillation in the $\overline{\nu}_e$ disappearance channel using 2\,200 days of data.  
We have obtained an 95\,\% C.L. excluded parameter region of $\sin^2 2\theta_{14}$  and $|\Delta m^2_{41}|$ for a mixing between $\overline{\nu}_e$ and a light sterile neutrino.
No evidence for a sub-eV sterile neutrino oscillation is found using two identical detectors, and thus yields a limit on $\sin^2 2\theta_{14}$ in $10^{-4} \lesssim |\Delta m^2_{41}| \lesssim 0.5$\,eV$^2$.
The RENO result provides the most stringent limits on sterile neutrino mixing at $|\Delta m^2_{41}| < 0.002$\,eV$^2$ using the $\overline{\nu}_e$ disappearance channel.

\vspace{2mm}
\noindent 
The RENO experiment is supported by the National Research Foundation of Korea (NRF) grants No.~2009-0083526, No.~2019R1A2C3004955, and 2017R1A2B4011200 funded by the Korea Ministry of Science and ICT. Some of us have been supported by a fund from the BK21 of NRF and Institute for Basic Science (IBS-R017-D1-2020-a00/IBS-R017-G1-2020-a00). We gratefully acknowledge the cooperation of the Hanbit Nuclear Power Site and the Korea Hydro \& Nuclear Power Co., Ltd. (KHNP). We thank KISTI for providing computing and network resources through GSDC, and all the technical and administrative people who greatly helped in making this experiment possible.
\bibliography{ref}

\end{document}